\documentclass[12pt]{article}

\usepackage[latin1]{inputenc}
\usepackage[english]{babel}
\usepackage{hyperref}
\usepackage{titling}

\newcommand{\be}{\begin{equation}}
\newcommand{\ee}{\end{equation}}

\usepackage{graphicx}
\usepackage{times}
\usepackage{fancyhdr}

\usepackage{epsfig}
\usepackage{multirow}

\usepackage{amssymb,amsfonts,amsmath}
\usepackage{color}

\begin{document}

\renewcommand{\multirowsetup}{\centering}

\title{Characterising two-pathogen competition in spatially structured environments}

\author{Chiara Poletto\textsuperscript{1,2,*}, Sandro Meloni\textsuperscript{3,4}, Ashleigh Van Metre\textsuperscript{1,2,5}, Vittoria Colizza\textsuperscript{1,2,6},\\ Yamir Moreno\textsuperscript{3,4,6}, Alessandro Vespignani\textsuperscript{6,7}\\}

\date{}

\maketitle

\begin{flushleft}
\small{\textsuperscript{1}Sorbonne Universit\'{e}s, UPMC Univ Paris 06, UMR-S 1136, Institut Pierre Louis d'Epid\'{e}miologie et de Sant\'{e} Publique, F-75013, Paris, France }\\
\small{\textsuperscript{2}INSERM, UMR-S 1136, Institut Pierre Louis d'Epid\'{e}miologie et de Sant\'{e} Publique,  F-75013, Paris, France }\\
\small{\textsuperscript{3}Institute for Biocomputation and Physics of Complex Systems, University of Zaragoza, Zaragoza, Spain}\\
\small{\textsuperscript{4}Department of Theoretical Physics, University of Zaragoza, Zaragoza, Spain}\\
\small{\textsuperscript{5}Wofford College, South Carolina, USA}\\
\small{\textsuperscript{6}ISI Foundation, Torino, Italy}\\
\small{\textsuperscript{7}Laboratory for the Modeling of Biological and Socio-technical Systems, Northeastern University, Boston MA, USA}\\

\vspace{1cm}
* Corresponding author: chiara.poletto@inserm.fr
\end{flushleft}

\newpage
{\centering \section*{Abstract}}

Different pathogens spreading in the same host population often generate complex co-circulation dynamics because of the many possible interactions between the pathogens and the host immune system, the host life cycle, and the space structure of the population. Here we focus on the competition between two acute infections and we address the role of host mobility and cross-immunity in shaping possible dominance/co-dominance regimes. Host mobility is modelled as a network of traveling flows connecting nodes of a metapopulation, and the two-pathogen dynamics is simulated with a stochastic mechanistic approach. Results depict a complex scenario where, according to the relation among the epidemiological parameters of the two pathogens, mobility can either be non-influential for the competition dynamics or play a critical role in selecting the dominant pathogen. The characterisation of the parameter space can be explained in terms of the trade-off between pathogen's spreading velocity and its ability to diffuse in a sparse environment. Variations in the cross-immunity level induce a transition between presence and absence of competition. The present study disentangles the role of the relevant biological and ecological factors in the competition dynamics, and  provides relevant insights into the spatial ecology of infectious diseases.

\newpage
\section*{Introduction}

The interaction between multiple infectious agents circulating within the same host population alters profoundly the spreading dynamics of infections and has important biological and public health implications~\cite{Keeling2008,Rohani2008}. Interaction mechanisms can have different nature and origins. Immune mediated interactions may affect polymorphic strains of a pathogen and represent a source of competition. This is the case of influenza A~\cite{Webster1992} in both humans~\cite{Zinder2013,Sonoguchi1985} and birds populations~\cite{Seo2001}, dengue in humans~\cite{Wearing2006}, foot and mouth disease in cattle~\cite{Haydon2001} and many others. In this case infection by a strain of the pathogen confers a certain level of immunity to other circulating variants. Immune mediated cooperation is observed as well. The antibody dependent enhancement in dengue represents a paradigmatic example, where cross-reactive antibodies following a previous infection increase the virulence of a subsequently infecting strain~\cite{Wearing2006}. Other examples include influenza versus Streptococcus pneumoniae~\cite{Opatowski2013}, and Malaria versus HIV~\cite{AbuRaddad2006}. Besides immunological mechanisms, ecological aspects can also represent a source of both competition and cooperation among pathogens. A permanent or a temporary depletion of hosts caused by a pathogen hampers the spread of another one, as in the case of measles and whooping cough~\cite{Rohani2003}, for example. Whereas an infection by Malaria is shown to increase individual's attractiveness to mosquitos~\cite{Lacroix2005} which in turn may increase chance to be infected by other strains. 

All these interaction phenomena 
are at the basis of pathogen evolution~\cite{Keeling2008,Roberts2011}. Despite the great interest in the problem, little is known on the drivers of the interaction dynamics and on the mechanisms ruling pathogens' ecological communities. A full understanding of the problem is hindered by the multiplicity and complexity of the mechanisms involved, from the microscopic scale of the interaction between the pathogen and the host immune system  to the global scale of hosts' behaviour and the environment~\cite{Pedersen2007}. 

At the population level, several modeling studies have addressed the problem in the context of pathogens evolution~\cite{Galvani2003, Gog2002, Koelle2006, Castillo-chavez89, Haraguchi2000, Boots2010,  Wild2009, Ballegooijen2004, Keeling2000, Webb2013}. These studies focus on multi-pathogen competition with the goal of understanding the evolutionary trade-off between transmissibility, infection duration and virulence. Some of them account for space structure~\cite{Haraguchi2000,  Boots2010,  Wild2009, Ballegooijen2004,  Keeling2000, Webb2013} in relation to the depletion of hosts induced by disease mortality~\cite{Haraguchi2000,Boots2010,Wild2009}, acquired immunity~\cite{Ballegooijen2004,Webb2013} and ability of the pathogen to persist in the population~\cite{Keeling2000}. The resulting picture, incorporating competition and evolution dynamics, is highly complex and often sensitive to the model's details. 

\begin{figure}[!b]
\begin{center}
\includegraphics[width=3in]{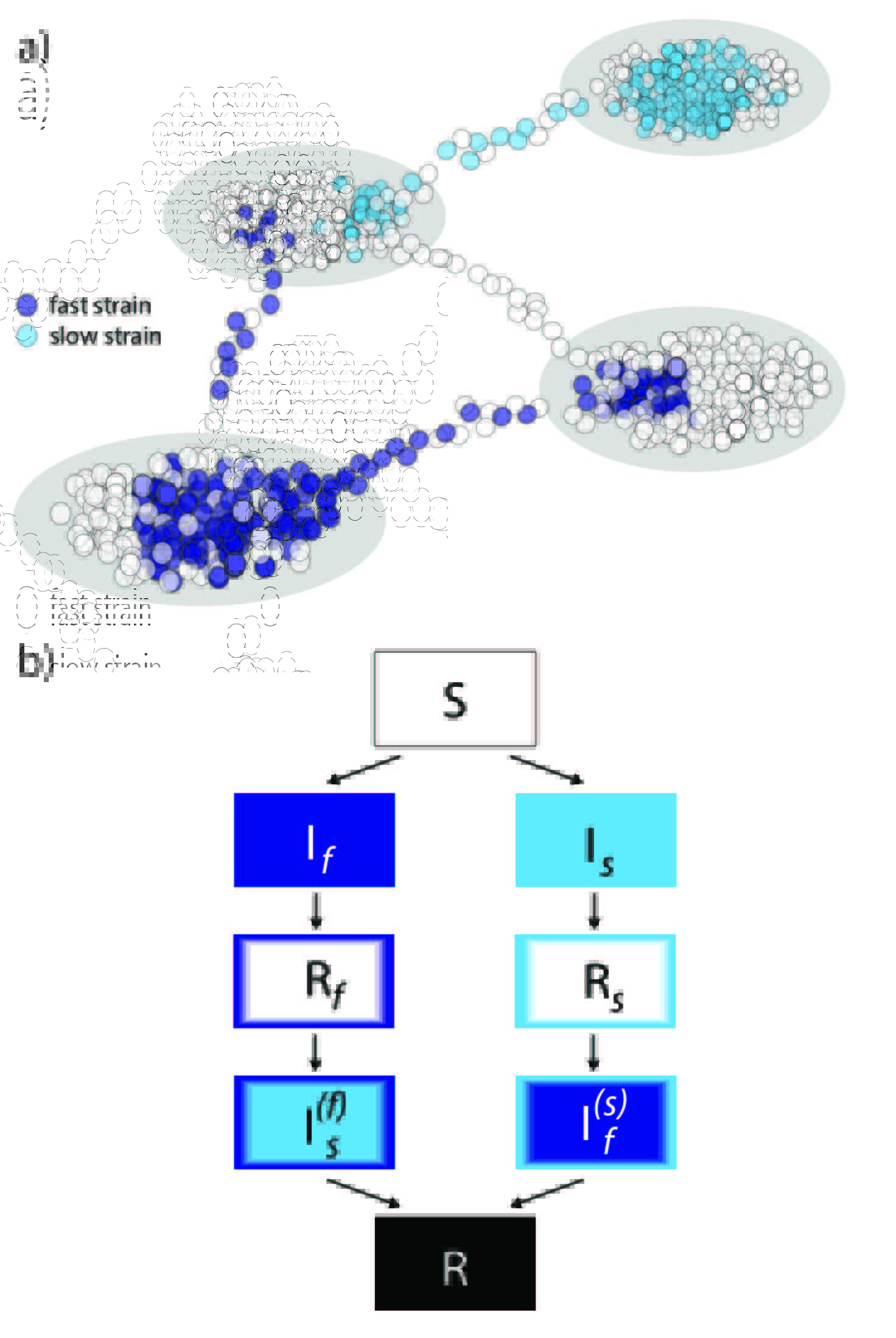}
\end{center}
\caption{Schematic representation of the metapopulation model with two pathogens. 
$a)$ Scheme of the metapopulation structure in patches and links representing mobility. $b)$ Compartmental model of the two-strain infection. A detailed description of the infection dynamics is reported in the Methods section.}
\label{fig:dia}
\end{figure}

Within this body of work little attention has been paid to the role of host mobility other than migration and recolonisation~\cite{Ballegooijen2004, Keeling2000}, an ingredient that potentially plays an important role in the case of rapidly spreading diseases for which depletion-replenishment considerations do not apply. Human population and many animal species (notably farmed animals) are characterised by complex mobility patterns that unfold at temporal timescales much faster than their life cycle~\cite{Bajardi2011b,Keeling2010,Chowell2003,Barrat2004,Gonzalez2008}. The network structure of such patterns and the traveling frequency have been shown to drive the spread of single-pathogen epidemics~\cite{Riley2007,Green2006,Keeling2005,Balcan2009a,Colizza2006a,Grais2004,Hufnagel2004,Merler2009,Grenfell2001}. In the context of two-pathogen competition induced by cross-immunity, an earlier study has shown that travel frequency determines the outcome of the competition in the case of full cross-immunity and when the competing pathogens have the same basic reproductive number, for infections conferring long-lasting immunity and during a single epidemic wave~\cite{Poletto2013a}. Here we build on this approach to explore systematically the role of epidemiological and immunological (i.e. reproductive numbers of the two pathogens and cross-immunity) and ecological parameters (spatial distribution of the hosts and mobility) in defining the co-circulation dynamics of competing pathogens. 
The competition dynamics is reconstructed through extensive  numerical simulations of a stochastic mechanistic model, and simple analytical considerations are found to explain the observed dynamics. The introduced modeling framework allows the characterization of the emerging competition dynamics and to describe the interplay of the different timescales of the processes involved. 
The simplifying assumptions considered here make the model applicable to a large variety of infectious diseases. Within this general framework we therefore use pathogen, strain, or variant as synonymous hereafter.

\section*{Results}

We consider two pathogens spreading in a spatially structured population of hosts modeled as a metapopulation system. The metapopulation modeling framework was originally introduced in population ecology~\cite{Levins69, Hanski2004} and later  applied to infectious diseases in order to account for a sparse distribution of hosts and consequently different levels of mixing~\cite{Anderson1984,Lloyd1996}. Several studies have recently coupled this framework with complex network approaches to account for non-trivial more realistic connectivity patterns among locations~\cite{Colizza2008, Colizza2007a, Balcan2011, Meloni2011, Poletto2013b, Liu2013, Apolloni2014, Belik2011a}. They assume individuals to mix homogeneously within the local communities (also called subpopulations, patches or nodes of the metapopulation network), whereas at the global level the coupling is defined by a network of hosts' mobility fluxes. Here we adopt this scheme considering two pathogens circulating on the metapopulation network (Figure ~\ref{fig:dia}a).
\begin{figure}[!]
\begin{center}
\includegraphics[width=1\columnwidth]{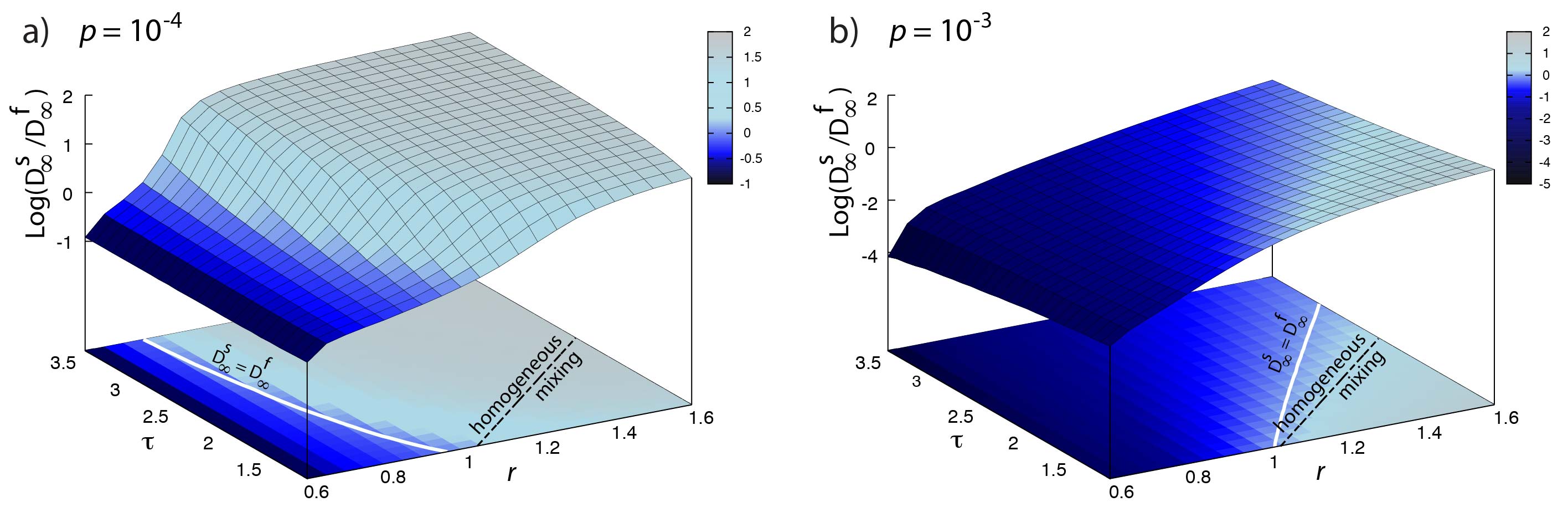}
\caption{\footnotesize{Competition between the two strains in the phase space $r,\tau$ for two distinct values of traveling probability $p$. The quantity in the $z$-axis is the logarithm of the ratio  $D_\infty^s/D_\infty^f$. Color code is proportional to the value in the $z$-axis, and the density plot in the horizontal plane shows the same quantity for the sake of visualisation. The white curve indicates the parameter region corresponding to the crossover where the two strains co-dominate, that is identified by $\log \left(D_\infty^s/D_\infty^f  \right)=0$. To highlight the effect the of mobility we report the crossover curve for the homogenous mixing case (black dashed curve). }} \label{fig:tau-r_p-fixed}
\end{center}
\end{figure}

We focus on the case in which the two pathogens confer long lasting immunity and interact through cross-immunity. To this end we follow  the multi-strain approach introduced by Castillo-Chavez et al. \cite{Castillo-chavez89} that assumes an individual recovered from one infection to have a susceptibility to the other circulating pathogen reduced by a factor $\sigma$ -- a schematic representation of the compartmental model is reported in Figure~\ref{fig:dia}$b$. The parameter $\sigma$ quantifies the level of cross-immunity, with $\sigma=0$ corresponding to complete cross-immunity and  $\sigma=1$ corresponding to no interaction. We allow the two circulating pathogens to have different transmissibility and recovery rates, indicated respectively with the parameters $\beta_{a}$ and $\mu_{a}$ for each pathogen $a$. Without loss of generality, we consider one of the two pathogens to have a slower infection progress, so that we can label the two pathogens {\it slow} and {\it fast} and introduce the parameter $\tau>1$ that quantifies the timescale separation,  $\mu_{s}^{-1} \equiv \tau \mu_{f}^{-1}$. With no interaction the impact of a pathogen in the population is fully determined by the basic reproductive number $R_0^{a} = \beta_{a}/\mu_{a}$ (with $a= \{ s,f \}$),  denoting the expected number of secondary infections generated by a single infectious individual in an entirely susceptible population and defining the condition for the epidemic to spread in a single population, i.e. $R_0^{a}>1$. We indicate with $r$ the ratio between the two basic reproductive numbers, $R_0^{s}= r R_0^{f}$. In the following we will explore in detail the role of the level of cross-immunity $\sigma$, and of the epidemiological differences between the two pathogens  encoded in the parameters $\tau$ and $r$. With this aim, we fix for simplicity $R_0^{f}$ and $\mu_{f}$ to realistic values for a generic acute infection~\cite{Keeling2008}, namely $R_0^{f}= 1.8$ and $\mu_{f}=0.6$ with a time unit of one day (corresponding to an infectious duration of $~1.7$days). The details of the compartmental classification are reported in the Methods section. Values explored lies in the range $\left[ 0.6,1.6\right]$ and $\left[1,3.5\right]$ for $r$ and $\tau$ respectively and the whole interval of definition $\left[0,1\right]$ for $\sigma$ -- see Table~\ref{tab:par}. While some combinations of parameters may be unrealistic  (e.g. antigenically similar strains inducing each other strong cross-protection are unlikely to have very different epidemiological traits) the goal of the present work is to provide a theoretical understanding of the dynamical behaviour of the system that is as general as possible and that can serve as ground of applied studies of diverse human and animal diseases. 

\begin{table*}[!ht]
\begin{tabular}{|l|l|l|}
\hline
{\bf Variable} & {\bf Description} & {\bf Values}\\
\hline
$V$ & number of patches & 10$^4$\\
$\bar N$ & average host population size per patch & $10^4$ \\
$k$ & patch degree, i.e. number of connections to other patches & average value $\bar k=5$\\
$p$ & travel probability & $\left[10^{-5},10^{-1}  \right]$\\
$R^f_0$ & reproductive number for the fast strain & $ 1.8 $\\
$\mu_f$ & recovery rate for the fast strain & $ 0.6 $\\
$R^s_0$ & reproductive number for the slow strain & $ R^s_0=rR^f_0$\\
$\mu_s$ & recovery rate for the slow strain & $ \mu_s^{-1}=\tau \mu_f^{-1} $\\
$\tau$ & timescale separation between the two pathogens & $\left[ 1,3.5\right]$ \\
$r$ & ratio of basic reproductive numbers of the two pathogens & $\left[ 0.6,1.6\right]$ \\
$\sigma$ & degree of cross-immunity & $\left[ 0,1\right]$\\
& & $\sigma=0$ full cross-immunity\\
& & $\sigma=1$ no interaction\\
\hline
\end{tabular}
\caption{ \label{tab:par} Model details and variables.}
\end{table*}

We consider a metapopulation with $V=10^4$ patches with an average population per patch $\bar N= 10^4$. Demography and mobility are modelled as follows. To each node $i$, we assign an initial number of individuals, $N_i$, and a degree $k_i$ denoting the number of connections the node has with other subpopulations in terms of mobility. Nodes' degrees  are distributed according to a Poisson probability distribution $P(k)$, which leads to a fairly homogenous topology and represents the simplest not trivial choice able to account for the small word property typical of empirical systems~\cite{Barrat2008,Erdos1959}. On top of the spatially structured system so defined, mobility fluxes are modelled by assigning to each individual in the subpopulation $i$ a probability $p$ per unit  time to travel to another neighbouring subpopulation $j$. We assume that departing individuals choose at random one of the available $k_i$ links \cite{Colizza2008}, so that the probability of traveling from $i$ to $j$ is given by $p/k_i$. According to the value of $p$, different mobility scenarios emerge: high values of $p$ yield large mobility fluxes resulting in a well mixed metapopulation system where individuals easily move from one patch to another; on the contrary small probability values correspond to  a dynamically fragmented scenario in which patches are fairly isolated. Different choices of network topology and fluxes distribution are clearly possible.  Comparison between homogenous and heterogenous topologies and fluxes distributions has been addressed before~\cite{Poletto2013a},  showing no qualitative differences in the observed dynamical behaviour among the different choices. On top of these structure, different kinds of mobility behaviour can be considered as well. For example, markovian displacements (i.e. movements where the memory of the origin of the traveling individuals is lost) are found in cattle and farmed animals, whereas origin-destination trips (thus highly non-markovian) characterize human travel. The impact of the mobility model on a single-pathogen epidemic spread has been extensively studied~\cite{Keeling2010, Balcan2011, Poletto2013b, Belik2011a}. Here we consider the simple case of markovian mobility to keep this part of the model as parsimonious as possible and better focus on the biological aspects.
A summary of the parameters and their values is reported in Table~\ref{tab:par}.

Once the system is initialised with a fully susceptible host population seeded with the two strains, the transmission dynamics of the two strains is reproduced by means of  Monte Carlo numerical simulations at the discrete host level. We consider hosts as integer units and we explicitly simulate both their mobility among different subpopulations and the infection transmission within each subpopulation as discrete-time stochastic processes. Throughout the analysis we will mainly consider as an indicator of the outcome of the competition between the two strains the final number of subpopulations $D_\infty^f$ and $D_\infty^s$ affected by each strain during the outbreak.

\begin{figure}[!]
\begin{center}
\includegraphics[width=1\columnwidth]{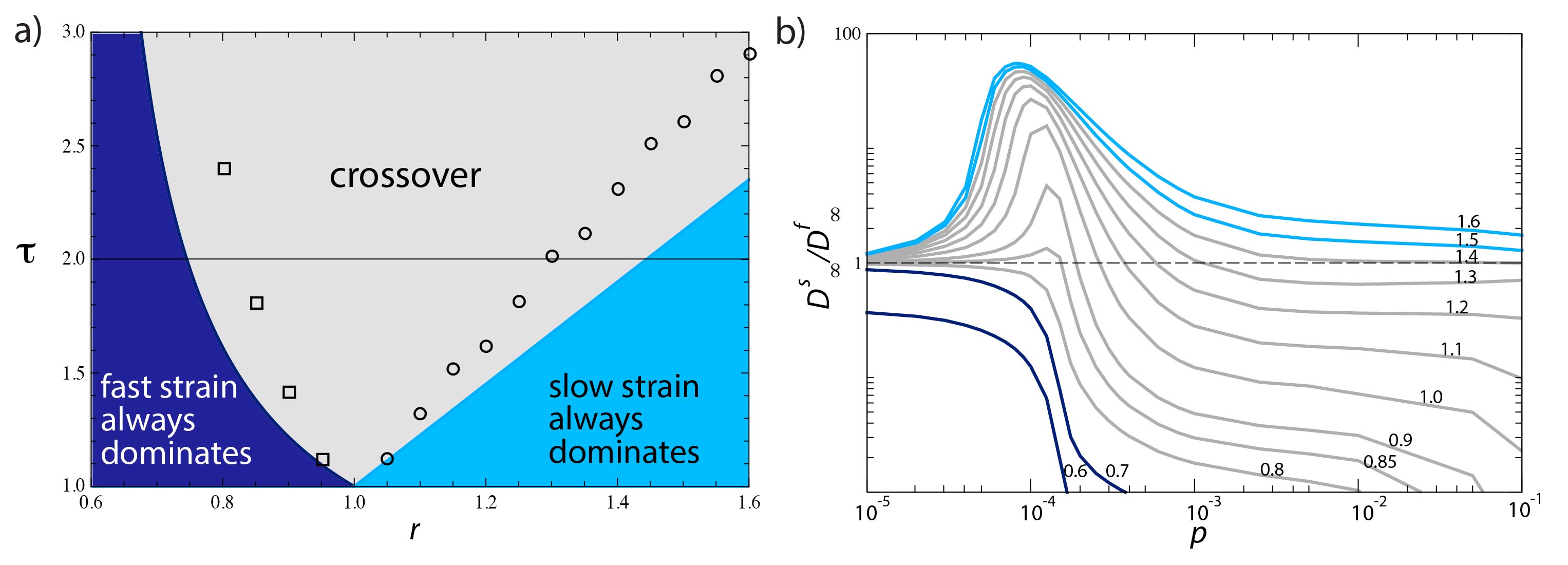}
\caption{\footnotesize{$a)$ Regions of presence or absence of crossover in the $(r,\tau)-$plane, as obtained by solving the inequalities of Eq.~(\ref{eq:crossover_cond}) for the case $R_0= 1.8$ and any $p$. Squares (circles) corresponds to the points for which crossover have been recovered in the numerical simulations in correspondence of $p=10^{-4}$ ($p=10^{-2}$). $b)$ Simulations of the two-strain spreading for the case $\tau=2$ (that corresponds to the slice of the left diagram indicated by the black line). The plot shows the ratio between the number of infected cities $D_\infty^s/D_\infty^f$ for different values of $r$. The colour code indicates the theoretical prediction: dark (light) blue curves correspond to the case where according to Eq.~(\ref{eq:crossover_cond}) the fast (slow) strain dominates for all values of $p$, while the grey curves are the ones for which the crossover takes place. The value of $r$ corresponding to each curve is indicated close to the curve itself.}} \label{fig:tau-r_phase-space}
\end{center}
\end{figure}

\subsection*{Signature of space in the two-pathogen competition}

We first consider the case of complete cross-immunity and analyse the impact of the difference in the epidemiological traits, encoded in the parameters $\tau$ and $r$, on the competition dynamics. In a homogeneous mixing population the pathogen with the largest growth rate $G=\mu (R_0 -1)$ dominates. Namely, it reaches more rapidly the majority of the population that cannot thus be infected by the other one~\cite{Keeling2008}. The relation $G_s= G_f$ defines then the boundary between the two  regions of the parameter space where either the slow or the fast strain is dominant. This translates into the linear relation 
\begin{equation}
rR_0-1=\tau(R_0 -1).
\label{eq:G_hm}
\end{equation}
Space structure and mobility completely alter this picture. Figure~\ref{fig:tau-r_p-fixed} shows the ratio $D_\infty^s/D_\infty^f$ for two different values of the traveling probability $p$. The light blue portion of the surface indicates the dominance of the slow strain, defined by the condition $D_\infty^s/D_\infty^f> 1$, whereas the remaining portion coloured in dark blue corresponds to the opposite situation. The boundary between the two regions (white curve in the surface projection on the $(\tau,r)-$plane in the Figure) changes under the different mobility regimes. The region of slow strain dominance is much larger in the reduced mobility scenario ($p=10^{-4}$) and, notably, the slow strain results to be the dominant even with a smaller transmission potential ($r<1$). This picture is qualitatively different from the homogenous mixing case that would divide the space of parameters according to Eq.~(\ref{eq:G_hm}), i.e. the black curve in the Figure. The scenario with higher $p$ is closer to the homogenous mixing case, as expected. The fast strain dominates over the slow one unless the latter does not have a considerable advantage in terms of basic reproductive number (i.e. $r$ large enough). The behaviour becomes increasingly closer to the homogenous mixing in the limit $p \to 1$.

Host mobility thus selects the epidemiological traits that are favoured in the multi-pathogen competition. A sparse environment would favour a slow pathogen, while a high overall mixing induced by large values of $p$ would favour the fast one. This result is analogous to the findings of previous studies focusing on the evolution of transmissibility and virulence, where virulence plays  a role analogous to the infection duration in the disease dynamics~\cite{Haraguchi2000,Boots2010}. In that case depletion and replenishment of individuals is the mechanism at the basis of  competition. A similar result has also been found in~\cite{Keeling2000}, where endemic diseases not conferring long lasting immunity were considered. In the present case the mechanism underlying the competition behaviour is the trade-off between the spreading velocity and the potential for spreading at the spatial level. The first is dictated by the timescale of the infection. The second quantifies the ability of the infection to generate a global epidemic by propagating out of the source through  infected traveling hosts. A full understanding of the competition diagram however requires to consider the invasion dynamic of the metapopulation system. 

\begin{figure}[!]
\begin{center}
\includegraphics[width=1\columnwidth]{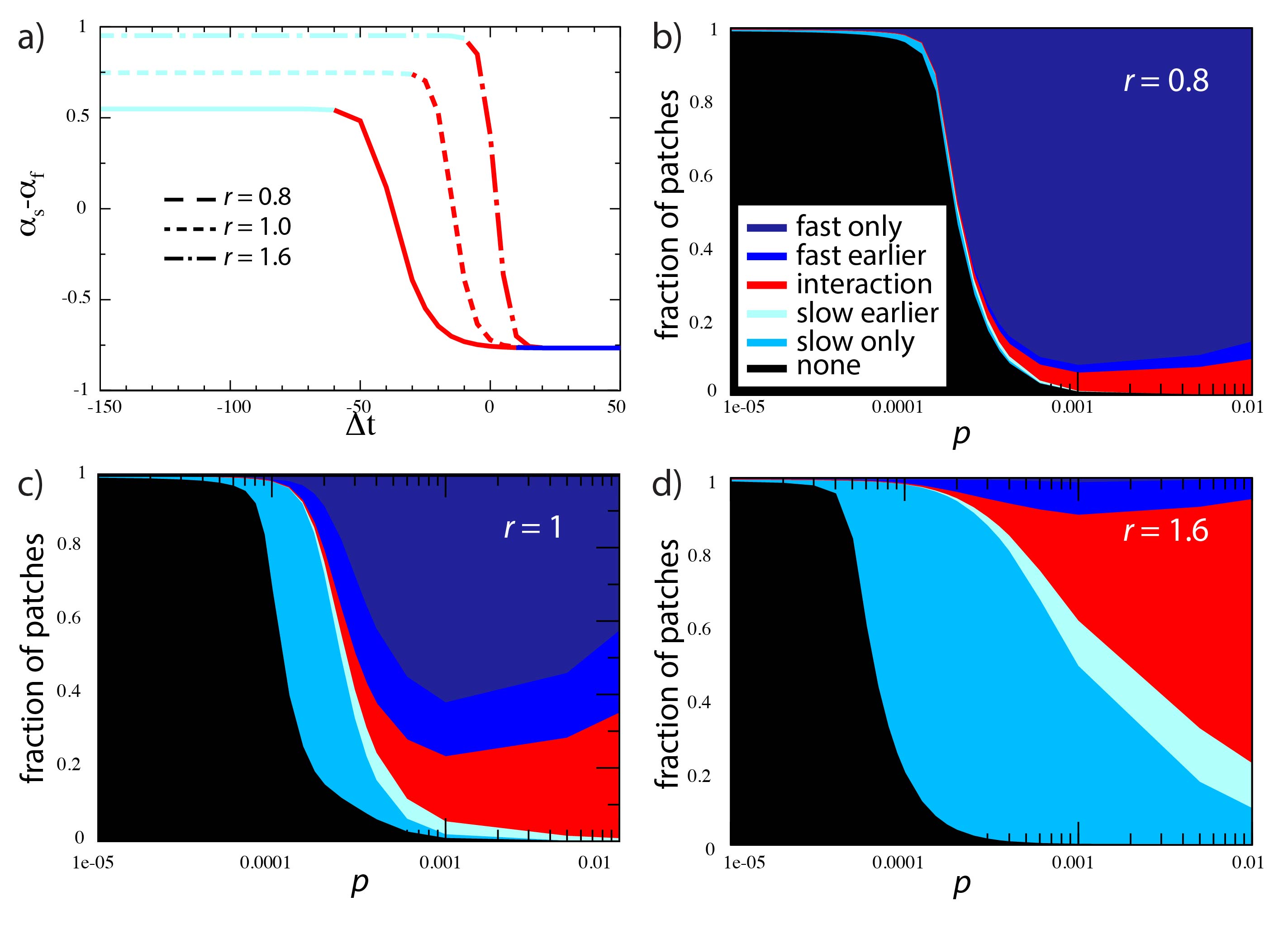}
\caption{\footnotesize{Focus on local co-circulation of the two strains within a patch. $a)$ two-strain competition within a single population as indicated by the difference between the attack rate of the slow and fast strains $\alpha_s-\alpha_f$, as a function of the delay in the seeding from the fast to the slow strain, $\Delta t= t_s-t_f$. Different value of $r$ are shown. The color code indicates different regimes where the fast strain spreads undisturbed (dark-blue), the slow strain reach the whole population before the fast epidemic is seeded (light-blue), the two seeding events are close in time in the way that the two pathogens interact (red part of the curve). The three panels $b)$, $c)$ and $d)$ summarise the situation at the metapopulation level as recovered by simulations. We measured the time of seeding of the two strains in the subpopulations and we classify the nodes according to their epidemic outcome that can be: $(1)$ the node is reached by the fast strain only; $(2)$ the node is reached by both strains with the fast one arriving early enough to spread undisturbed in the population; $(3)$ the epidemic is seeded by both strains  with small delay, thus allowing interaction; $(4)$  the epidemic is seeded by both strains with a timing corresponding to the slow strain dominance; $(5)$ the node is reached by the slow strain only; $(6)$ the node is reached by none of the strains. Three distinct values of $r$ are compared; $\tau=2$.}} \label{fig:delay}
\end{center}
\end{figure}

\subsection*{Conditions for spatial induced crossover in the competition} \label{sec:cross-over_condition}

In order to rationalize the role of the metapopulation structure in the outcome of the two-pathogen competition observed in Figure~\ref{fig:tau-r_p-fixed}, we consider the observable $R_*$, an  
additional predictor introduced to synthetically describe the conditions for the spatial invasion and that accounts for all the biological and behavioural mechanisms involved in the spatial spread (i.e. pathogen traits, host traveling behavior, structure of the mobility network and demography). The global invasion parameter $R_*$  defines the invasion threshold $R_*>1$~\cite{Poletto2013a,Colizza2008,Cross2007}. Analogously to $R_0$ at the individual level, $R_*$ represents the average number of patches an infected  patch is able to infect before the end of the local outbreak. If this is greater than one a global outbreak will occur, otherwise the epidemic will be confined around the initially seeded subpopulation. For the case of a homogenous mobility network, the analytical expression of the invasion threshold is quite simple~\cite{Colizza2008}. This allows us to frame analytically the trade-off between spreading velocity, upper-bounded by the exponential growth in the homogenous mixing case ($G$), and the spatial invasion potential, encoded in $R_*$. We assume all nodes to have the same degree $\bar k$. For each pathogen $a$ the functional dependence of $R^a_*$ on the variables of the system is given by~\cite{Poletto2013a,Colizza2008}:
\begin{equation}
R^a_*= \left(\bar k-1\right)\left(  1-\left(  \frac{1}{R^{a}_{0}}  \right)^{  \frac{p  {\boldsymbol \alpha_a}  \bar N}{\mu_a \bar k}  } \right),
\label{eq:thresh_hom}
\end{equation} 
where  $\lambda_{ij}=\frac{p  {\boldsymbol \alpha_a}  \bar N}{\mu_a \bar k}$ is the average number of infectious individuals that travel from an infected patch, $i$, to a neighbouring one, $j$, during the entire duration of the outbreak in $i$~\cite{Colizza2008}. This factor  depends  on the traveling probability $p$, on the epidemiological parameters through the attack rate ${\boldsymbol \alpha_a}=\alpha_a(R_0^a)$ (resulting from the circulation of the strain $a$ only in the fully susceptible population) and $\mu_a$, and on demographic ($\bar N$) and spatial ($\bar k$) features. The expression $1-\left ( \frac{1}{R^{a}_0}\right )^{\lambda_{ij}}$ is the probability that an outbreak is seeded in $j$ by the $\lambda_{ij}$ infectious travelers~\cite{Bailey1975}. The factor $(\bar k -1)$ represents the number of connections along which the disease can spread (all possible connections $\bar k$ except the one where the infection comes from). 

The pathogen that is more efficient in propagating at the spatial level is the one with higher $R^{a}_{*}$ and  infectious period. The effect of the infectious period is due to the fact that if an infected individual stays infectious longer has more chance to travel while ill and bring the disease in new patches, in the simple SIR formulation under study. This means that the slow strain is favoured, provided that the fast one does not have a much larger reproductive number. The balance between these factors becomes critical in the regime of small traveling probabilities. The observable $R^{a}_{*}$ is an increasing function of $p$, thus for large enough $p$  $R^{a}_{*} \gg 1 $ for both strains indicating that they both spread easily through the system. In this regime the  ingredient determining the competition outcome is the spreading velocity that favours, for comparable $R_0$, the strain with short infectious duration. As a consequence of this trade-off, mobility can induce a crossover between the two regimes of fast and slow strain dominance. 

On the basis of the considerations above the condition for the crossover can be easily recovered. Given that in the limit $p \to 1$ the exponential growth of the pathogen $a$ is given by $G^{a}$, the crossover is encountered whenever 
\begin{eqnarray}
G^{f}>  G^{s} \;\ \; \text {and} \; \; R^{s}_{*} > R^{f}_{*}.
\label{eq:crossover_cond}
\end{eqnarray}
We notice however that when $r>1$ the inequality $R^{s}_{*} > R^{f}_{*}$ is always satisfied and, on the other hand, the condition $G^{f}>  G^{s}$ is guaranteed when $r<1$. Therefore the condition above can be rewritten explicitly in the following  form:  
\begin{eqnarray}
& r>1 &  G^{f}>  G^{s} \Rightarrow \; r R_0^f -1 < \tau (R_0^f-1), \nonumber  \\
& r< 1&   R^{s}_{*} > R^{f}_{*}  \Rightarrow \; \tau {\boldsymbol \alpha_s} \log(R_0^s) > {\boldsymbol \alpha_f}  \log(R_0^f).
\label{eq:crossover_cond}
\end{eqnarray}
Network and demographic parameters factorise in the above expressions. This analytical reasoning is simplified and is based on the assumption that the two epidemics do not interact, which is strictly verified only in the limit of infinite network size.

The left side of Figure~\ref{fig:tau-r_phase-space} displays the phase-space of the presence/absence of crossover for the case $R_0^f= 1.8$ as obtained by Eq.~(\ref{eq:crossover_cond}). For all the $\left(r,\tau \right)$ values in the grey region a crossover takes place at a certain value of $p$. In reference to the examples reported in Figure~\ref{fig:tau-r_p-fixed}, we provide the crossover curves obtained for $p=10^{-4}$ (squares) and $p=10^{-3}$ (circles), corresponding to the white curves of the previous Figure.
The right panel of Figure~\ref{fig:tau-r_phase-space} further describes the crossover behaviour by displaying the average ratio $D_\infty^s/D_\infty^f$ obtained from  simulations as a function of $p$ and for different values of $r$. In order to facilitate the comparison with the theoretical results we colour coded the curves according to the expected competition outcome. Results show a good agreement between simulations and analytical reasoning despite the simplicity of the latter, indicating that the previous theoretical argument is able to capture the fundamental mechanisms underlying the competition between the two strains. According to the relation among the epidemiological traits of the two pathogens, the level of  coupling induced by mobility may be either a determinant factor or a non-influential one for the competition outcome. 

Strains inducing strong cross protection each other are likely to be highly genetically similar and to have small difference in their epidemiological traits, i.e. $r$ and $\tau$ close to one. In this case the unbalance between spreading velocity and/or potential for spatial spread would be small and so also the advantage of one strain over the other. However even under these conditions the system preserves its rich behaviour. Figure~\ref{fig:tau-r_phase-space}a indeed shows that all three competition regimes are included in the neighbourhood of the point $\left(r,\tau \right)=(1,1)$ of the space of parameters.

\subsection*{Local interaction and overlap between the epidemic waves}

The outcome of the competition presented in the previous section is the result of the spreading pattern of the two pathogens and their  interaction. At the local level of a single patch, this is certainly due to the conditions of arrival of each pathogen seeding the local population. In this subsection we intend to address the local co-existence between the two pathogens, given its important epidemiological implications, for example for pathogen recombination~\cite{Lindstrom2004}.  

For a given set of parameters $r$ and $\tau$ the outcome of the competition within each single patch depends on the time delay between the seeding of the two epidemics, $\Delta t= t_{s}-t_{f}$, where $t_{a}$ is the time of seeding of the strain $a$ in the patch. 
Figure~\ref{fig:delay}a characterises the epidemic impact through the observable $\alpha_s-\alpha_f$, i.e. the difference between the attack rates produced by the slow and  fast strains when they are co-circulating. In the case in which the fast strain arrives first, the slow strain has minimal chance to reach a significant fraction of the population. For positive values of $\Delta t$, the quantity $\alpha_s-\alpha_f$ rapidly approaches the limit value ${\boldsymbol \alpha_f }$ obtained in absence of strain $s$ (dark blue portion of the plot). Referring to the case $r=1$ as an example, if the population is seeded by the slow strain with more than 10 days delay with respect to the fast strain, the slow strain is prevented from spreading in the population. It can dominate over the fast one only in the case in which it has a significant advantage in terms of time of arrival (or seeding).  The difference  $\alpha_s-\alpha_f$ saturates to the limit value of ${\boldsymbol \alpha_s } $ only if the slow epidemic starts 40 days in advance with respect to the fast one (light-blue portion) for $r=1$. The two pathogens co-circulate and interact at the population level within the patch only if they seed the population with a small delay one with respect to the other (red portion). This condition defines the interaction time window, whose length depends on all the parameters of the system. The larger the value of $r$, the faster the two epidemics spread, thus increasingly reducing the time-window for possible interactions within a patch.

At the metapopulation level such phenomena occur at the interface of the propagation waves of the two epidemics, i.e. the set of nodes that are reached by both strains. This interface region can be limited to few nodes or can be almost as large as the whole network according to the values of traveling probability and the disease parameters. However, for the interaction to occur, the two epidemics need to be seeded on such interface region with a delay within the interaction time window, otherwise we observe a complete separation of the two propagation phenomena (i.e. one strain swiping the population much after the first strain has already circulated). The extent of the interface region where interaction occurs increases with increasing mobility and  is strongly affected by $r$ (Figure~\ref{fig:delay}).
Large values of $r$ make local interaction between pathogens more likely as shown by the red portion of the plot that becomes wider as $r$ increases. This occurs despite the range of $\Delta t$  is narrowed by large values of $r$ and it is due to the fact that larger reproductive numbers in the slow strain counterbalance the shorter infection period yielding a similar exponential growth to the fast pathogen. This results in a larger region of intersection between the two epidemics.

\begin{figure}[!]
\begin{center}
\includegraphics[width=\columnwidth]{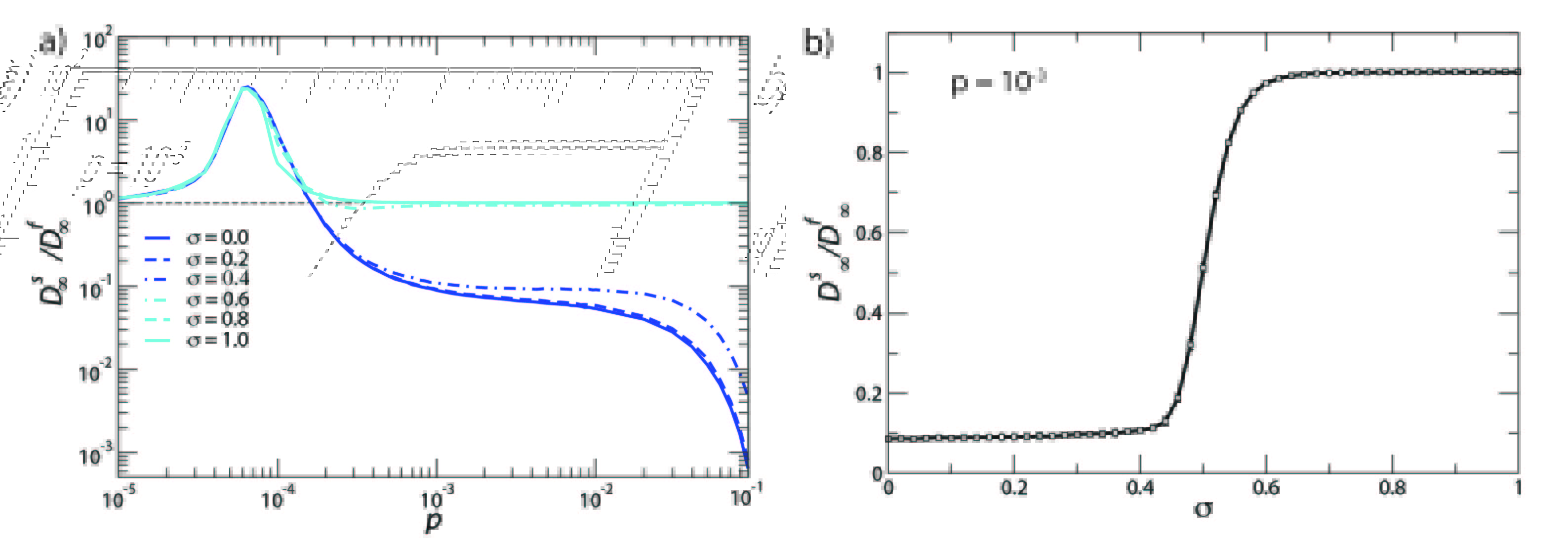}
\caption{\footnotesize{Role of partial cross-immunity. $a)$ Simulation results  of the two-strain spreading for the case $\tau=2$. The plot shows the ratio $D_\infty^s/D_\infty^f$ as function of $p$ for different values of $\sigma$. $b)$ the plot shows the ratio $D_\infty^s/D_\infty^f$  as a function of $\sigma$ for $p=10^{-3}$.}} 
\label{fig:cutsigma}
\end{center}
\end{figure}

\subsection*{Role of partial cross-immunity}

Two interacting pathogens or strains of the same pathogen  are rarely antigenicaly equal, so that the level of cross-immunity is  smaller than 1. An intermediate level of cross-immunity indicates that individuals recovering from one strain are partially susceptible to the other~\cite{Smith2004}. This can be accounted for in our compartmental model by considering $\sigma>0$. 

We address how this ingredient affects the competition dynamics at the spatial level by first considering the case $r=1$. Figure~\ref{fig:cutsigma}a shows the ratio $D_{\infty}^{s}/D_{\infty}^{f}$ for different values of the cross-immunity parameter $\sigma$. Varying $\sigma$ from full cross-immunity ($\sigma=0$) to no cross-immunity ($\sigma=1$) we observe a sharp transition between two distinct classes of behaviour. For $\sigma$ below a given threshold, the small level of susceptibility after the first infection does not impact the competition dynamics and the curves strictly follow the full cross-immunity case with a crossover for intermediate mobilities and the dominance of the fast strain, $D_\infty^{s} \ll D_\infty^{f} $, for $p \to 1$. The behaviour changes rapidly for $\sigma$ above a critical value $\sigma_c \simeq 0.4$.  All the curves for $\sigma>\sigma_c$ display the same behaviour with the two pathogens co-circulating and reaching the same portion of the network ($D_\infty^{s}/D_\infty^{f} \simeq 1$) for all values of $p$ except for a range of small values of $p$ where $D_\infty^s>D_\infty^f$. Figure~\ref{fig:cutsigma}b displays the sharp transition behaviour in $\sigma$ obtained for $p=10^{-3}$.

The result can be explained in a simple way in terms of  herd immunity effects induced by the partial cross-immunity~\cite{Keeling2008}. In a well mixed population the first strain to reach the population (e.g. strain $a$) infects a proportion ${\boldsymbol \alpha_a}$ of individuals who have a reduced probability to be infected by the second strain (strain $b$) after their recovery from the infection. The spread of strain $b$ is then ruled by the effective  reproductive number that results from the combination of its transmission potential and the level of susceptibility of the population, $R^{\text{eff}}= R_0^a \left( 1- {\boldsymbol \alpha_a}  +\sigma\,{\boldsymbol \alpha_a}  \right)$. The condition $R^{\text{eff}}>1$ allows then the second pathogen to generate an outbreak in the subpopulation and further spread in the neighbouring patches. If applied to the scenario discussed above, for large values of $p$ where both pathogens have the potential to spread at the spatial level and  competition is determined by the spreading velocity, all levels of cross-immunity such that $R^{\text{eff}}<1$ are equivalent in suppressing the spread of the slower pathogen. For $\sigma$ rising above the threshold, this mechanism is not anymore in place and both strains propagate through the system reaching the whole network. Even in this case, however, the two strains do interact, as measured by the reduction of the attack rate of the slow strain. The mechanism favouring the spread of the slow strain for small values of $p$ is still present since it is not  due to the interaction between the two pathogens but to the larger invasion potential $R_*^s$ of the slow strain with respect to the fast one.

If $r\ne 1$, i.e. the two pathogens have different basic reproductive numbers, the emerging picture is more complicated. For small mobility rates, variations in the degree of cross-immunity do not lead to quantitative variations in the competition outcomes (expect for values $\sigma \simeq 1$), and the ratio $r$ rules the dynamics. For higher values of $p$, instead, both cross-immunity and the relative ratio of the reproductive numbers of the two pathogens determine the resulting competition outcome. We report these results in the Supplementary Figure S1. 

\begin{figure}[!]	
\begin{center}
\includegraphics[width=0.5\columnwidth]{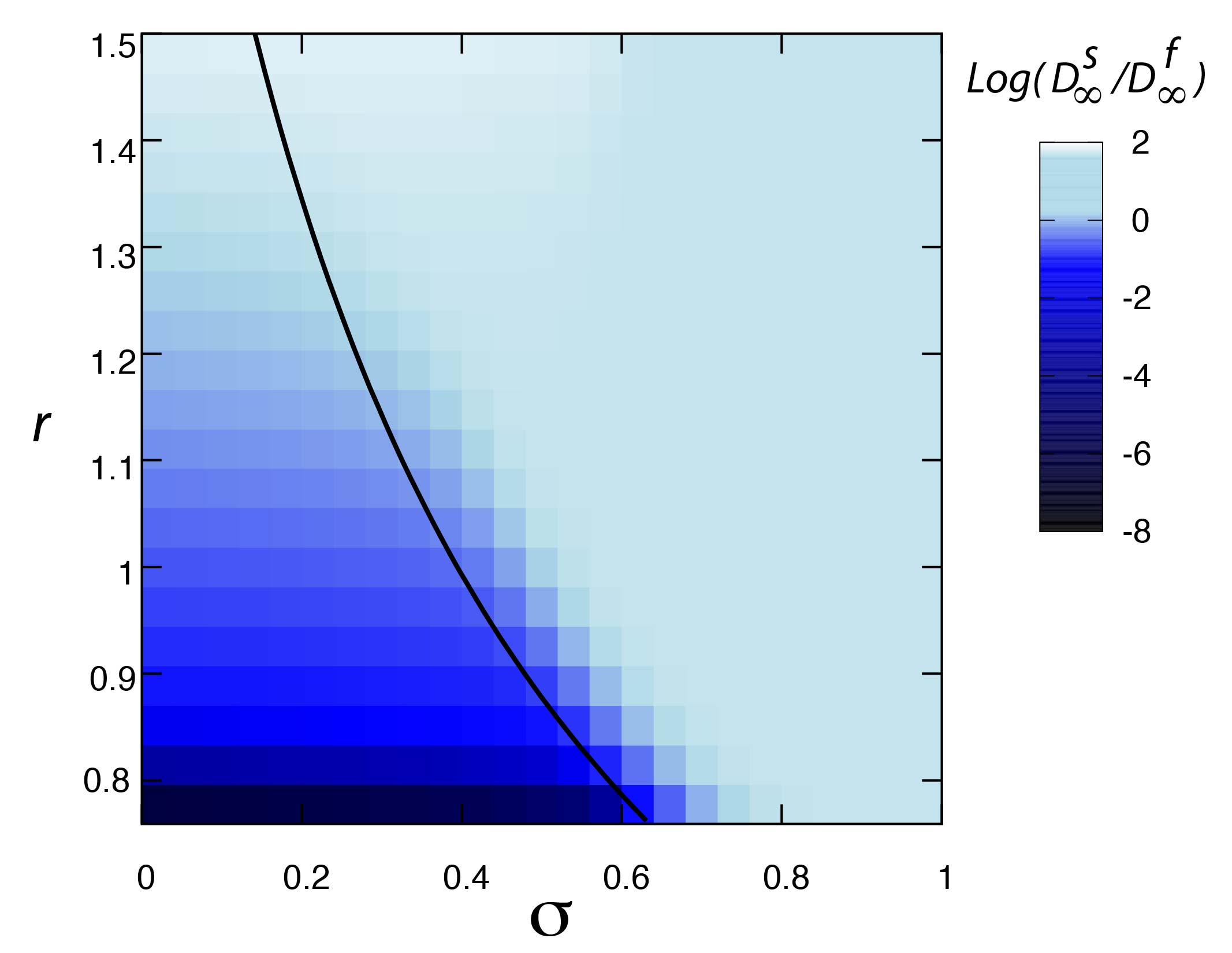}
\caption{\footnotesize{density plot of the ratio  $D_\infty^s/D_\infty^f$ in the $(r,\sigma)-$plane for travelling probability $p=0.01$. The black curve indicates the condition $R^{\text{eff}}_0= r R_0 ( 1-\alpha(R_0) +\sigma\,\alpha(R_0)) =1$. Simulations are obtained with $\tau=2$.}} \label{fig:heatmap}
\end{center}
\end{figure}

It is worth noting that for very high values of the mobility rate, when competition is ruled exclusively by the epidemic growth rate, the argument based on the herd immunity effects is able to explain the observed results~\cite{Keeling2008}. When $R^{\text{eff}}= r R_0^f ( 1-{\boldsymbol \alpha_f} +\sigma\,{\boldsymbol \alpha_f}) >1$ the strain with smaller exponential growth loses its disadvantage and becomes able to spread in the metapopulation system. This is shown in the density plot in Figure~\ref{fig:heatmap} where  $p=0.01$ is considered. Despite the complexity of the spatial propagation dynamics recovered in the stochastic simulations, very simple analytical reasonings are able to shed light on the mechanisms underlying the spreading. 

\section*{Conclusion}

We studied the role of host mobility in the immune-mediated competition between two pathogens causing acute infections. We provided an extensive numerical characterisation of the interaction dynamics by varying the degree of cross-immunity and the difference in the epidemiological traits of the two pathogens (basic reproductive numbers and infectious periods). Depending on the relation between the pathogens' traits, mobility can play a determinant role or be non-influential for the outcome of the competition. In the space of epidemiological parameters, there exists a region for which lowering the traveling probability induces a cross over from the fast strain dominance to the slow strain one. This behaviour is determined by the trade-off between epidemic growth and potential for spread at the spatial level, the former being an advantage in a well mixed population while the second being relevant in a sparse environment with intermediate or low mobility coupling. 

This non trivial result has important implications for disease ecology. The role of hosts' space structure in the multi-pathogen competition and the impact of changes in hosts' mixing patterns on the disease evolution still represent open research questions~\cite{Haraguchi2000, Boots2010, Ballegooijen2004, Webb2013}. Our study characterises the epidemiological conditions under which hosts' traveling behaviour is a determinant ingredient in the competition dynamics. Reducing the degree of cross-immunity determines a rapid transition between the picture described above to a situation of no competition in the spatial propagation. This transition behaviour can be framed within a herd immunity paradigm, where the more rapid pathogen acts as a vaccine in the spreading dynamics of the other one. 

As a consequence of the sharp transition in $\sigma$, results described in Figures~\ref{fig:tau-r_p-fixed} and \ref{fig:tau-r_phase-space}, remain substantially unchanged in the more realistic case of partial cross-immunity, considering a wide range of values of $\sigma$ (e.g. $\sigma \lesssim 0.4$, with the parametrisation adopted here) compatible with epidemiological estimates (see for example~\cite{Zhang2013} for influenza). It is important to note, that a certain level of short-term cross-protection due to immunological mechanisms other than memory-cell immunity has been observed among antigenically distant strains (e.g. in the case of dengue~\cite{Wearing2006}, different types and subtypes of human and avian influenza~\cite{Sonoguchi1985,Seo2001}, hemorrhagic disease in wildlife \cite{Park2013}), and it is indicated in many cases as the source of competition during a time scale of a single epidemic wave~\cite{Wearing2006,Sonoguchi1985,Seo2001}. In this case, a certain degree of heterogeneity in the transmissibility and/or infectious duration can be reasonably assumed and in some cases it is also documented. Epidemiological evidence includes influenza A subtypes in birds~\cite{Saenz2012}, dengue serotypes~\cite{Wearing2006}, and pandemic vs. seasonal subtypes in human influenza~\cite{Fraser2009} (possibly as a consequence of a higher level of immunity in the population to the seasonal strain). All these multi-strains diseases represent examples of  dynamical systems for the infection propagation whose study could  benefit by the theoretical understanding here provided.

The modelling framework here introduced allows us to account  for important features characterising host mobility patterns in a realistic way. Despite the complexity of the dynamics simulated by the mechanistic model, the analytical formulation of the global invasion potential allows for simple analytical considerations able to shed light on the behaviour observed in numerical simulations. The network formalism is an important ingredient of the model and represents an element of novelty with respect to previous studies on disease ecology where the space is introduced by placing individuals on a regular lattice~\cite{Haraguchi2000, Boots2010, Ballegooijen2004, Webb2013} or by assuming a metapopulation with two levels of mixing~\cite{Wild2009, Keeling2000} (high mixing within patches and low mixing between patches). Other studies adopt network approaches~\cite{Funk2010,Newman2011} for studying multi-pathogen competition, however they do not account for different levels of mixing  or for host traveling. 

The framework here introduced has the potential to provide an important understanding of the multi-pathogen dynamics in more realistic and complex situations. Additional important mechanisms that were not considered in this study are worth to be mentioned. The competition was analysed during a single epidemic wave, given that no demographic turn-over nor waning of immunity were considered. These factors are crucial for many diseases, e.g. human and avian influenza and dengue.  The two epidemics were assumed to start at the same time in a fully susceptible population.  Emergence events where a new pathogen or a new variant starts spreading in a population already affected by other seasonally circulating variants represent however a source of great concern. Influenza again provides a paradigmatic example with its frequent zoonotic events yielding new virus subtypes that represent a threat for human population (e.g. A-H1N1pdm09~\cite{H1N1team}, A-H7N9~\cite{Gao2013}). Given the general nature of the metapopulation model considered, our approach proposes a theoretical and computational framework where these additional ingredients can be further considered and implemented to deepen our understanding of pathogens competition.

\section*{Methods}

\subsection*{Infection dynamics}

The infection is modelled through the compartmental scheme of Figure~\ref{fig:dia}. Individuals are divided in susceptible ($S$), infected by the slow strain with no previous infection history ($I_s$), infected by the fast strain with no previous infection history ($I_f$), recovered by the slow strain ($R_s$), recovered by the fast strain ($R_f$),  infected by the slow strain previously infected by the fast one ($I^{(f)}_s$), infected by the fast strain previously infected by the slow one ($I^{(s)}_f$), permanently recovered and immune to both strains ($R$). Susceptible individuals can contract the infection by either the fast or the slow pathogen with probability $\beta_f \left(I_f + I^{(s)}_f  \right)/N$ and $\beta_s \left(I_s + I^{(f)}_s  \right)/N$ respectively, with $\beta_{a}=R^{a}_0/\mu_{a}$ being the transmission rate of pathogen $(a)$ and $N$ the population of the patch. Individuals recovered by one pathogen can contract the infection by the other one with the same probability as above reduced by a factor $0 \leq \sigma \leq 1$. Regardless of the infection history, infected individuals recover with the pathogen specific recovery rate, i.e. $\mu_s$ and $\mu_f$ for the slow and fast pathogen respectively. In the stochastic mechanistic simulations contagion and recovery are modelled as binomial and multinomial processes. The step of the simulation $\delta t$ defines the unitary timescale of the process and corresponds to one day. 

\subsection*{Mobility network and traveling}

The mobility network is generated following the Erd\H{o}s-R\'{e}nyi algorithm~\cite{Erdos1959}, which consists of assigning a link between each pair of nodes with probability $\bar k/(V - 1)$. This results in a Poisson degree distribution with fairly homogenous topology. Specifically the network considered in the study has average degree $\bar k= 5$, $V= 10^4$ nodes and diameter equal to $8$. The traveling of hosts is implemented by assuming each individual to travel with probability $p$. This mobility process yields a population distribution at the equilibrium given by
\begin{equation}
 N_i= k_i \bar N / \bar k,
 \label{eq:eqpop}
 \end{equation}
where $\bar N$ is the average population size. 

In the stochastic mechanistic simulations the traveling is implemented as follows. For each subpopulation $i$, the number of traveling individuals are extracted from each of the eight infectious compartments through a multinomial distribution characterised by $k_i+1$ possible outcomes which correspond to traveling to each of the $k_i$ directions, with probability $\frac{p}{k_i}\delta t$, and to not traveling, with probability $1-p \, \delta t$. $\delta t$ is the same used for the infection dynamics.

\subsection*{Computational modeling of competing pathogens}

To simulate the spread of the two strains on the metapopulation system of susceptible hosts, we initialise the number of individuals of each subpopulation at the equilibrium value given by Eq.~(\ref{eq:eqpop}).  This guarantees the system to be at the equilibrium of the mobility dynamics in such a way that the population of each node fluctuates around the initial value for the whole duration of the simulated outbreak without any significant replenishment/depletion of individuals. The epidemic is initialised  by seeding 50 randomly extracted subpopulations with the slow (fast) strain and moving  0.1\% of the population to the $I_s$ ($I_f$) compartment, keeping the rest of the population in the susceptible compartment. We explicitly required that the two strains are not initialised within the same nodes to avoid interaction at the beginning of the epidemic. We tested different number of initially infected subpopulations (i.e. $10$ and $25$) obtaining the same qualitative results. 

For each set of parameters we simulate 2,000 stochastic realisations of the spatial epidemic spreading randomly selecting different initial conditions and different instances of the mobility network. Traveling across patches and infection transmission within each patch are simulated at every time step until the epidemic gets extinct, i.e. until  the $I_s$, $I_f$, $I^{(s)}_f$ and $I^{(f)}_s$ compartments are empty in all subpopulations.  For each run we record the attack rate within each subpopulation produced by both the fast and the slow strain, as well as the time of arrival of each strain. In calculating the number of infected subpopulations by each strain, $D_\infty^f$ and $D_\infty^s$, we consider that a patch is infected by a strain if at least a fraction $\alpha_T$ of the population within the patch has contracted the disease. We set $\alpha_T$ equal to 10\% and we checked that the results are not sensitive to the value of this parameter. Quantities displayed in the plots are averages over all  runs.

\section*{Acknowledgements}
This work has been partially supported by the EC-Health contract no. 278433 (PREDEMICS) to VC; the ANR contract no. ANR- 12-MONU-0018 (HARMSFLU) to VC; MINECO through Grant FIS2011-25167 to YM; Comunidad de Arag\'on (Spain) through a grant to the group FENOL to YM;  the EC Proactive project MULTIPLEX (contract no. 317532) to SM, YM and AlVe. AsVa has been supported by the Papadopoulos Scholarship at Wofford College. AlVe has been partially funded by the DTRA-1-091003 and NSF CMMI-1125095 awards. SM is supported by the MINECO through the Juan de la Cierva Program.

\section*{Author contributions}
Conceived and designed the experiments: CP SM VC YM AlVe. Performed the experiments: CP SM AsVa. Analyzed the data: CP SM AsVa. Wrote the paper: CP SM VC YM AlVe. Developed the model: CP SM VC YM AlVe.

\section*{Additional Information}
The authors declare no competing financial interests.

\end{document}